\documentclass[showpacs,aps]{revtex4}
\usepackage{graphicx}

\begin{document}

\title{  Baryonic hybrids:  Gluons as beads on strings between quarks}
\author{John M. Cornwall\footnote{Email:  Cornwall@physics.ucla.edu}}
\affiliation{Department of Physics and Astronomy, University of California, Los
Angeles CA 90095
\begin{center}
{\rm (Received December xx, 2004)}
\end{center}}

\begin{abstract}
\pacs{11.15.-q, 12.38.-t, 11.15.Tk   \hfill UCLA/04/TEP/51}
  
 In this paper we
analyze  the ground state of the heavy-quark  $qqqG$ system using standard
principles of quark confinement and massive constituent gluons as established
in  the center-vortex picture.  The known string tension $K_F$ and
approximately-known  gluon mass $M$   lead to a
precise specification
of the long-range non-relativistic part of the potential binding the gluon to
the quarks with no undetermined phenomenological parameters, in the limit of
large interquark separation $R$.    Our major tool (also used
earlier by Simonov) is the use of proper-time
methods to describe  gluon propagation within the quark system, along with some
elementary group theory   describing the gluon Wilson line as a composite of
co-located $q$ and $\bar{q}$ lines.  We show that (aside from color-Coulomb
and similar terms) the gluon potential energy in the presence of quarks is
accurately described (for small gluon fluctuations) via attaching these three
strings to the gluon, which in equilibrium sits at the Steiner point of the
Y-shaped string network joining the three quarks.   The gluon undergoes small
harmonic fluctuations  that slightly stretch these strings and quasi-confine
the gluon to the neighborhood of the Steiner point.  To describe
non-relativistic ground-state gluonic fluctuations at large $R$
we use the Schr\"odinger equation, ignoring mixing with $l=2$ states. 
Available lattice data and real-world
hybrids require consideration of $R$ values small enough for significant
relativistic corrections, which we apply using a variational principle for the
relativistic harmonic oscillator.  We also consider the role of  color-Coulomb
contributions.   In terms of interquark
separations $R$, we find  leading non-relativistic large-$R$ terms in the
gluon excitation energy  of the form 
$\varepsilon (R)\rightarrow M+\xi [K_F/(MR)]^{1/2}-\zeta \alpha_c/R $
where $\xi ,\zeta $ are  calculable numerical coefficients and $\alpha_c\simeq
$ 0.15 is the color-Coulomb $q\bar{q}$ coupling.  When the gluon is
relativistic, $\varepsilon \sim (K_F/R)^{1/3}$. We
get an acceptable fit to lattice data with $M$ = 500 MeV.  Although we do not
consider it in full detail, we show that in the $q\bar{q}G$ hybrid the gluon is
a bead that can slide without friction on  a string joining the $q$ and
$\bar{q}$.   We comment briefly on the
significance of our findings to fluctuations of the minimal surface, a subject
difficult to understand from the point of view of center vortices.
\end{abstract}

\maketitle

\section{Introduction}

In QCD, hybrids are hadrons that cannot be understood as combinations of quarks
and antiquarks only, but that require consideration of gluonic excitations
along with the quarks.  Perhaps the most interesting of these are the exotic
hybrids,
which are hadrons whose quantum numbers are not found in any colorless
combination of $q$ and $\bar{q}$, and so must contain valence gluons. 
Hybrids and exotics \cite{jaffe} have been studied theoretically for decades
(for example,
\cite{corntuan}, \cite{isgpaton}), but these early studies are tied to models
and approximations that hardly make it possible to choose one model from
another. Experiments have searched for hybrids as well, and there are a number
of credible observed candidates for mesonic hybrids (for a
recent review  see \cite{meyer}).  The author knows of no plausible candidates
for baryonic hybrids. 

Unfortunately, theoretical approaches to
hybrids are, naturally enough, even more forbidding than approaches to
ordinary hadrons.  So one might wonder what the value of a theoretical paper
on hybrids is.  The answer is that recent lattice simulations \cite{tsuga,jkm}
of heavy static quarks (described by straight Wilson lines in the Euclidean
time direction) with single-gluon excitations provide a valuable
testing-ground for theoretical ideas, because one does not have to deal with
the extra complication of quark Wilson-line fluctuations and because lattice
data are available at theoretically-interesting separations of quarks.

We will be primarily interested in estimating the
ground-state gluonic excitation energy $\varepsilon (R)$ of the heavy-quark
baryonic hybrid $qqqG$, as a function of a suitably-defined interquark
separation $R$, and in comparing our results to lattice data \cite{tsuga}.   
We will also remark in less detail on similar issues for mesonic hybrids, but
will not attempt to compare to data in this paper.  The regime where we are on
firmest ground is the regime of asymptotically-large large $R$, where the
dynamics is non-relativistic (because the gluon has a dynamical mass) and we
can fairly confidently use the Schr\"odinger equation.  However, lattice data
\cite{tsuga} are available for smaller $R$, where relativistic corrections are
important, and to cover the entire relevant range of $R$ we use a variational
principle incorporating fully-relativistic kinetic energy for the gluon.  The
smallness parameter for non-relativistic motion is a numerical constant times
$[K_F/(M^3R)]^{1/2}$, where $K_F$ is the string tension
and $R$ the interquark separation.  It turns out that our calculations are in
excellent agreement with the lattice data for a gluon mass of 500 MeV, even
for $R$ small enough for relativity to matter.  

One should note that in the fully-relativistic
regime there should be mixing with other states containing more gluons, such
as baryon plus glueball.  Additionally, the derived potentials have angular
momentum $l=2$ terms as well as $l=0$.  These lead to mixing of gluonic $l=2$
and $l=0$ states, and should lower the energy slightly compared to the results
we give.   This lowering might be compensated by raising the gluon mass.  We
do not consider either of these mixings here.

The  idea that the gluon has a constituent mass $M$ \cite{corn82} goes back a
long way, but a purely theoretical determination of $M$ is not yet within our
powers.   Lattice simulations have verified this idea and found values for the
mass, although many such simulations, done in particular gauges, cannot
guarantee a gauge-invariant determination of the mass.  However, Ref.
\cite{deforc} has extrapolated their Euclidean lattice data to the Minkowski
regime, to find the so-called pole mass, which is gauge-invariant.  
The pole mass so found on the lattice  is
$M\simeq$ 600 (+150 -30) MeV \cite{deforc}.    We believe that our fitted value
of 500 MeV is acceptably close to the lattice pole mass,  given the expected 
uncertainties in our approach and in the lattice determinations.

 There are two main steps in finding the dependence of the gluonic energy on
$R$.  The first uses a proper-time technique for describing
propagation of the gluon in the hybrid, and the second, as described above, 
implements the dynamical 
fluctuations in the proper-time gluon paths with a Schr\"odinger equation
modified for relativistic effects.  In the first step, we describe the
propagator of the gluon in the heavy-quark
hybrid as a quantum perturbation on a background of confining gauge
potentials. The  center-vortex picture of confinenment prescribes
\cite{corn04,corn041} that the quarks are joined by a Y-shaped network of
minimal surfaces (or strings, at fixed time).  The gluon propagator has a
proper-time
form that is an integral over all paths of an adjoint Wilson line going along
a path joined at both ends to the system of heavy-quark Wilson lines. [These
techniques were developed earlier by Simonov \cite{simon} and applied to
$q\bar{q}G$ hybrids, but not $qqqG$ hybrids.  Simonov's works were unknown to
the present author, who developed them independently and discovered Simonov's
work after this paper was finished.]  As we will see, the result is that if we
do not distinguish the three quarks by flavor or other quantum numbers, the
gluon can be described as a massive bead attached to the three strings joining
the quarks in the hybrid, and fluctuating in directions transverse to the
string. The fluctuations stretch the string slightly, and there is a harmonic
restoring force on the gluonic bead.  In the $q\bar{q}G$ hybrid the gluon is a
bead sliding without friction on the string joining the quark and antiquark.

All the potentials we use in the second step are harmonic, found by expansion
of various minimal areas occurring in the baryonic Wilson loop. Except in the
ultra-relativistic limit and the Newtonian limit
this relativistic Schr\"odinger equation is not analytically solvable, but good
approximations to $\varepsilon (R)$ for large $R$ can be
found by a variational technique.  We describe this technique
in the Appendix.  For the $qqqG$ system, the color-Coulomb energy can be
straightforwardly estimated classically and added as a perturbation.  It is
not quite so straightforward for the $q\bar{q}G$ hybrid color-Coulomb energy,
which we describe in the Appendix.  The Appendix also treats the string energy
for the $q\bar{q}G$ hybrid, finding relativistic gluon energies  whose leading
term is of the form $N\pi /R$.  Lattice data on heavy-quark $q\bar{q}G$
systems suggest such leading string-like energies, but with substantial
corrections \cite{jkm,juge}.  These corrections can be estimated with our
techniques, but we do not study the problem here.

Our final result yields three terms in the large-$R$ (non-relativistic)
expansion of the $qqqG$
energy:
\begin{equation}
\label{larger}
\varepsilon (R) \simeq M+\xi (\frac{K_F}{MR})^{1/2}-\frac{\zeta \alpha_c}{R}
+\cdots
\end{equation}
where $\xi ,\zeta$ are  numerical parameters  that we estimate with fair
accuracy below and in the Appendix, and $\alpha_c$ is the coefficient of $-1/R$
in the $q\bar{q}$ Coulomb energy.  
At shorter distances, where relativistic corrections are important, the gluon
energy scales like $(K_F/R)^{1/3}$, and Coulomb corrections are smeared out by
the gluon wave function, as discussed in the Appendix.   
 
 There are numerous other corrections to the small-$R$ results, including
spin-dependent terms and poorly-understood terms that arise \cite{corn04} when
two Wilson
loops partly coincide, as is the case for quark Wilson loops formed from the
gluon propagator.  We ignore all such terms, which do not seem to be very
large, judged by our fit to the lattice data.  It is somewhat surprising that
our techniques give good results even for fairly short distances, considering
all the possible effects that can enter.

\section{Proper-time techniques and Wilson loops}
\label{propertime}

 In gauge-theory dynamics, a Wilson line or loop is to be integrated not only
over all gauge configurations but also over all possible lines joining the
fixed endpoints, with a certain weight function.  This weight function encodes
all the particulars about the mass, flavor, and spin of the particle being
described by the Wilson line.  In other words, particle propagation is to be
described by a proper-time path integral.  

\subsection{Gluon proper-time propagator}

We wish to construct the necessary path integral for a gluon in a hybrid state,
where distance scales are large compared to the QCD scale.  We will argue
below that gluonic spin contributes only short-range effects, which allows us 
to simplify matters by considering the propagation of a scalar particle in the
adjoint representation.  In a fixed background gauge potential the Euclidean
proper-time propagator $\Delta (x; y)$ from $x$ to $y$ of a scalar particle of
mass $M$ is
\begin{equation}
\label{proptime}
\Delta (x; y)= {\mathcal{N}}\int_0^{\infty}\int_x^y(dz)\exp
\{\frac{-M}{2}\int_0^s d\tau [\dot{z}^2+1]\}U_A(x; y)
\end{equation} 
where $(dz)$ stands for the integral over all paths $z_{\mu} (\tau )$ from $x$
at $\tau$=0 to $y$ at $\tau =s$; ${\mathcal{N}}$ is a normalization factor;
and $U_A(x;y)$ is the adjoint Wilson line
\begin{equation}
\label{adjointline}
U_A(x;y)=P\exp [ig\int d\tau 
\dot{z}_{\mu}T^{\alpha}A_{\mu}^{\alpha}(z)].
\end{equation} 
Here $T^{\alpha}$ are the adjoint generators, and $P$ stands for path-ordering.
Of course, the propagator $\Delta$ is not gauge-invariant by itself, but we
will soon couple it to quarks in a gauge-invariant way.

If (still ignoring spin) we identify $\Delta$ as the propagator of the
potential $A_{\mu}^{\alpha}(z)$ that occurs in $U_A$, intractable problems
arise.  So we envision the gauge potential as being separated into a confining
part (condensate of center vortices) that constitutes the background potential
and the small-amplitude part whose
propagation we are describing; it is this latter part that constitutes the
gluonic excitation in a hybrid hadron.  As for spin itself, the propagation of
a vector gauge boson with small amplitude in a background field requires that
$\dot{z}_{\mu}$ in $U_A$ be replaced by $\dot{z}_{\mu}+\Sigma_{\mu
\nu}\partial_{\nu}$, where $\Sigma_{\mu \nu}\partial_{\nu}$ are the spin-one
generators of rotations in Euclidean four-space, and path ordering is also
carried out with respect to the spin operator.  However, the derivative yields
the background potential field strength, which is short-ranged.  Since our
primary interest is in long-range effects, we will from now on ignore gluon
spin, and think in terms of the explicit scalar propagator of Eq.
(\ref{proptime}).
 
\subsection{Splitting the adjoint Wilson line}

It is familiar fact that a gluon line in a Feynman diagram can
be replaced by co-located quark and antiquark lines, joined to other quarks
according to certain rules. The usual large-$N$ counting rules, for example,
follow from such considerations.   Perhaps less familiar is the dynamical
application of this splitting of a gluon line \cite{simon}, which we now
describe.  The resulting picture is most easily described for $q\bar{q}G$
hybrids, which we discuss first.

\subsubsection{Mesonic hybrids}

\begin{figure}
\includegraphics[height=3in]{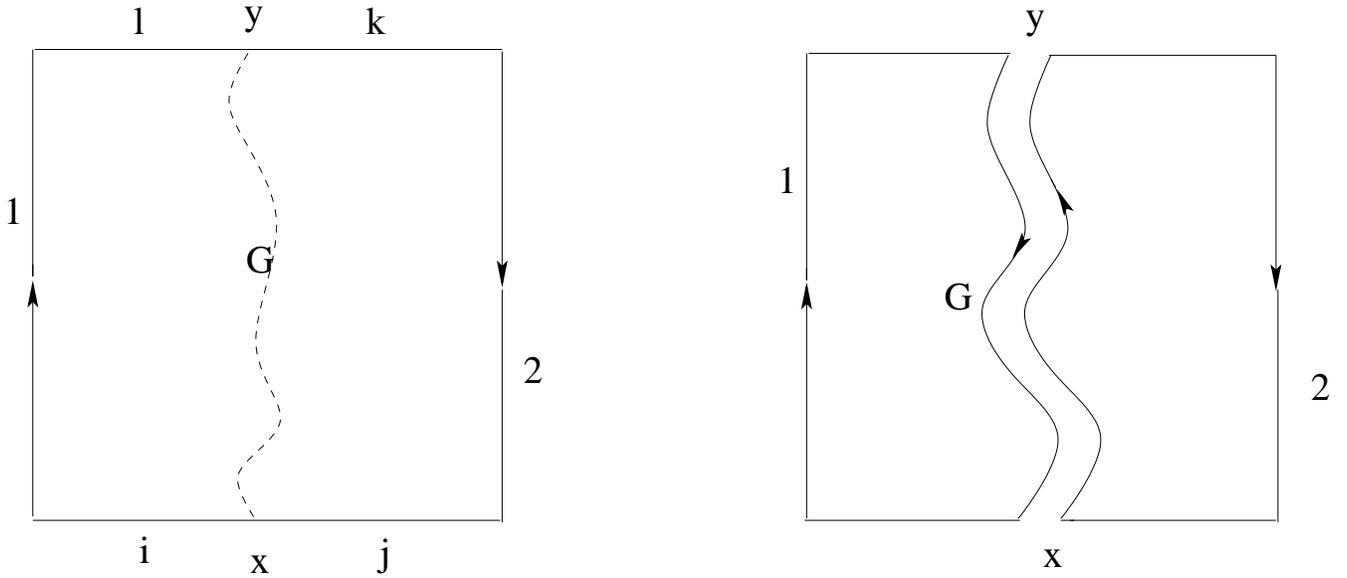}
\caption{\label{qqgloop} Left:  A rectangular Wilson loop for a quark (1) and
antiquark (2) and a gluon (G).  Right:  The same loop with the gluon line
decomposed into another quark and antiquark line, forming two standard Wilson
loops.}
\end{figure}

Let
\begin{equation}
\label{hybridop}
H_{\nu}(x)=\bar{\psi }\gamma_{\mu}\frac{1}{2}\lambda^{\alpha}\psi 
G_{\mu\nu}^{\alpha}(x)
\end{equation}
be the interpolating field for a hybrid meson, with $(1/2)\lambda_{\alpha}$ the
conventionally-normalized generator in the fundamental representation.  (It
happens to be the field describing an exotic, with $J^{PC}=1^{-+}$
\cite{corntuan}.)  The field $\psi $ describes very massive quarks.  Ignoring
spin indices and other irrelevant complications, the $H$-field propagator
$\Delta_H$ is an expectation value
\begin{equation}
\label{hfield}
\Delta_H = \langle U^i_l(1)U^k_j(2)[U_A(g)]^{jl}_{ik}\rangle
\end{equation}
where $U(1, 2)$ are fundamental Wilson lines for the quark and antiquark, and
$\langle \cdot \rangle$ means to integrate over all gauge potentials and over
the gluon Wilson line as shown in Eq. (\ref{proptime}).  The gluon Wilson line
matrix has been converted into a matrix with fundamental indices on it, via
\begin{equation}
\label{fundadj}
[U_A(g)]^{jl}_{ik}=\frac{1}{2}(\lambda_{\alpha})^j_i(\lambda_{\beta})^l_k
[U_A(g)]_{\alpha\beta}.
\end{equation}
Here $U_A$ is the adjoint representative of the group element $g$, which is the
path-ordered product seen in Eq. (\ref{adjointline}).  We show the index
structure of Eqs. (\ref{hfield}, \ref{fundadj}) in Fig. \ref{qqgloop}.

Now we split the gluon line, using the $SU(N)$ equations
\begin{equation}
\label{lambda}
\frac{1}{2}(\lambda_{\alpha})^j_i(\lambda_{\beta})^l_k=\delta^j_k\delta^l_i-
\frac{1}{N}\delta^j_i\delta^l_k
\end{equation}
and
\begin{equation}
\label{adjfund}
[U_A(g)]_{\alpha\beta}=\frac{1}{2}Tr[U(g)\lambda_{\alpha}U^{\dagger}(g) 
\lambda_{\beta}]. 
\end{equation}
Combining leads to a familiar equation:
\begin{equation}
\label{combine}
[U_A(g)]^{jl}_{ik}=U^l_i(g)[U^{\dagger}(g)]^j_k-\frac{1}{N}\delta^j_i\delta^l_k.
\end{equation}
This equation has the group property: If $g=g_1g_2$, the product
\begin{equation}
\label{product}
[U_A(g_1)]^{ja}_{ib}[U_A(g_2)]^{bl}_{ak},
\end{equation} 
when computed using the right-hand side of Eq. (\ref{combine}), has the form of
that right-hand side with $g$ replaced by $g_1g_2$.  

For us, $g$ is the group element in the gluon Wilson line [see Eq.
(\ref{adjointline})]. The adjoint on the right-hand side of Eq.
(\ref{combine}) means to trace the path backwards, as one easily checks.
Therefore the hybrid propagator in Eq. (\ref{hfield}) assumes the simple form
\begin{equation}
\label{hsplit}
\Delta_H(x;y)=\langle TrU(1G)TrU(G2)\rangle +\dots
\end{equation}
where 1$G$ is the closed fundamental Wilson loop shown in the right side of
Fig. \ref{qqgloop}, and similarly for the other closed loop $G$2.  The omitted
term comes from the $1/N$ term in Eq. (\ref{combine}), and yields a single
Wilson loop describing a $q\bar{q}$ pair.  We will drop this term, which in
any case cannot contribute for exotic hybrids.

\subsubsection{Baryonic hybrids}

Similar considerations apply for $qqqG$ hybrids, but there are minor
group-theoretic complications. The direct product of three quark
representations is
\begin{equation}
\label{dirprod}
3\otimes 3\otimes 3 = 1 \oplus 8 \oplus 8' \oplus 10
\end{equation} 
There are two 8s which can couple to a gluon. Denote the quark group vectors by
$a^i,b^k,c^l$, and form the octet combination 
\begin{equation}
\label{qqqv}
V^i_j(a;b,c)=\epsilon_{jkl}a^ib^kc^l -\frac{1}{3}\delta^i_j\epsilon_{pqr}
a^pb^qc^r.
\end{equation} 
Quark $a$ has been singled out (its index is external).  By permuting the quark
group vectors one seems to come to three possible octets, but because of
\begin{equation}
\label{vident}
V^i_j(a;b,c)+V^i_j(b;a,c)+V^i_j(c;a,b)\equiv 0
\end{equation}
there are really only two.  

\begin{figure}
\includegraphics[height=3in]{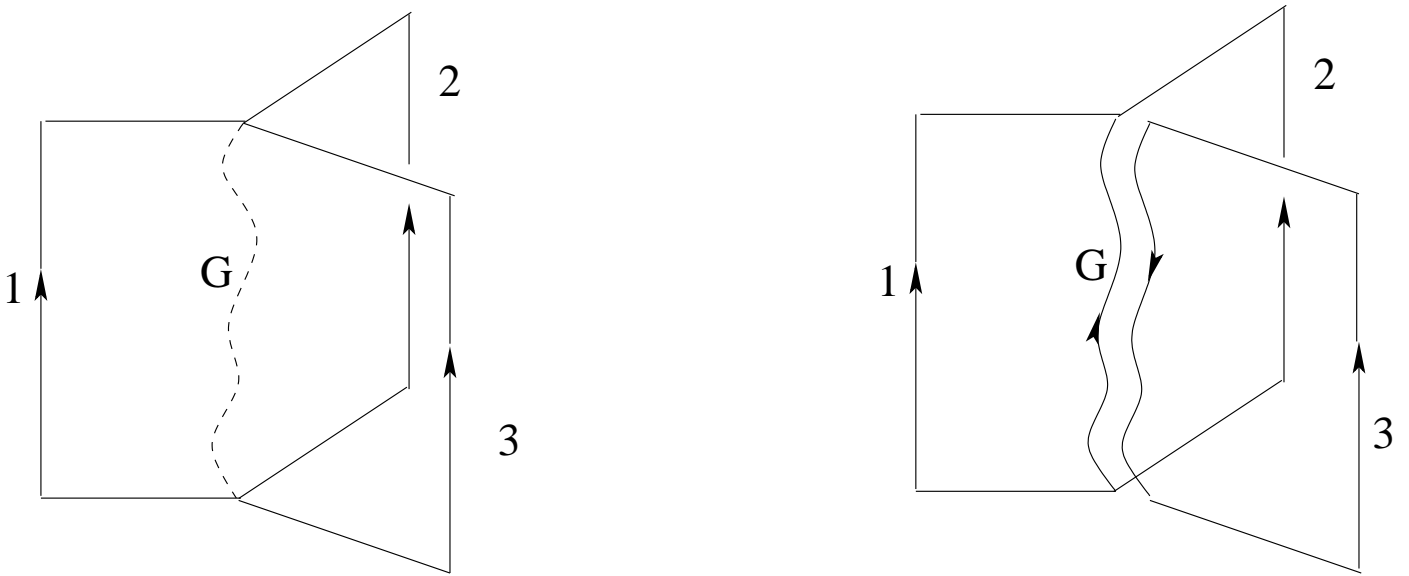}
\caption{\label{qqqloop} Left:  Wilson loop for a $qqqG$ configuration (quarks
labeled 1, 2, 3; gluon labeled G).  Right:  The same loop with the gluon line
decomposed into another quark and antiquark line, with quark line 3 singled
out as discussed in the text.}
\end{figure}

We can construct a propagator for a $qqqG$ hybrid in the same spirit as the
earlier work for the mesonic hybrid.  Let there be three heavy-quark lines
labeled 1, 2, and 3, plus gluon line $G$.   We write the propagator for one of
the two octets, singling out line 3 as we did quark $a$ above (again dropping
the trace term, which is missing the gluon):
\begin{equation}
\label{qqqgloop}
\langle
\epsilon_{abc}\epsilon^{pqr}U^a_p(1)U^b_q(2)U^i_j(3)U_A(g)^{cj}_{ir}\rangle .
\end{equation}
Using the identity of Eq. (\ref{combine}) as before yields the $qqqG$ analog of
the gluon line splitting for mesons as shown graphically in Fig.
\ref{qqqloop}.  Line 3, which has been singled out, forms a $q\bar{q}$ state
with one of the split gluon lines, while the other split gluon line forms a
new baryon 12$G$.

\subsection{Gluon dynamics}

In the quark Wilson loops of Figs. \ref{qqgloop}, \ref{qqqloop}, the
rectangular lines do not fluctuate (the quarks are infinitely heavy).  But the
lines associated with the gluon do fluctuate, since the gluon path-integral
weight function has a finite mass.  The dynamics of the quark loops is
governed by conventional area laws, and we now turn to the question of how
those area laws govern the gluon dynamics.

The center-vortex picture of the area laws for a fundamental Wilson loop claims
\cite{corn04} that, without the extra gluonic effects of the present paper,
the area to be used in the area law is the minimal area spanning the given
Wilson loop.  (This only makes sense if there is a unique such area, which is
certainly true in the present case; see \cite{corn04} for further discussion.) 

\subsection{Qualitative behavior of gluonic potentials} 
 
\begin{figure}
\includegraphics[height=3in]{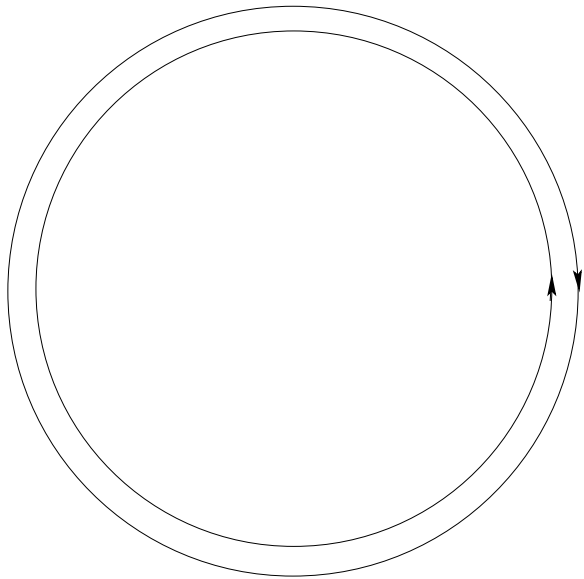}
\caption{\label{gluonloop} An adjoint Wilson loop displayed as the
superposition of two oppositely-oriented Wilson loops.}
\end{figure}

Let us begin with some qualitative remarks.  Consider a Wilson loop in the
adjoint representation, which we
decompose into two fundamental (quark) Wilson loops in the now-familiar way as
shown in Fig. \ref{gluonloop}.  Because the superposed loops are
oppositely-oriented, the linkage of a center vortex with one loop is exactly
and coherently cancelled by the linkage of the same vortex to the other
(oppositely-oriented) loop.  Thus the center-vortex linkage contributes a
factor of unity to the adjoint Wilson loop, as must be the case.

Consider the right-hand part of Fig. \ref{qqgloop}, showing the gluon world
line decomposed as a quark and antiquark line.  The area law for the original
$q\bar{q}$ pair, labeled 1 and 2, is the sum of the two areas on the right
(which of course is the original area on the left side of the figure), as long
as the gluon line stays in the original minimal surface (plane of the figure)
and does not stray too close to the lines 1 and 2.  (As the gluon line
approaches these lines,  perimeter terms arise because one or the other of the
two Wilson loops in the right-hand part of the figure has a length scale
comparable to the QCD scale).  That is, fluctuations of the gluon line which
are confined to the original minimal surface change nothing; the gluon does
not feel the $q\bar{q}$ pair.  

\begin{figure}
\includegraphics[height=3in]{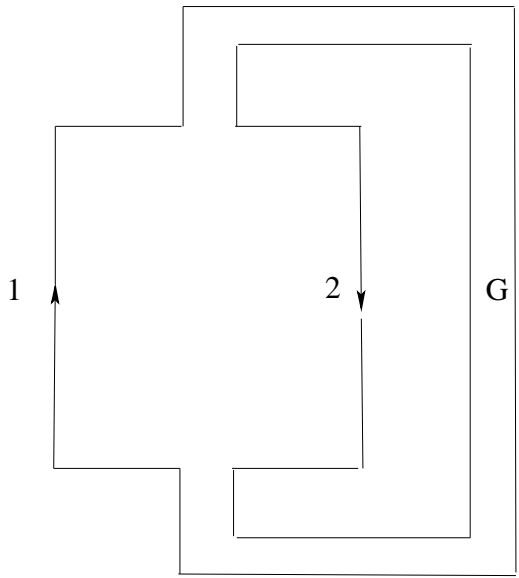}
\caption{\label{qqloop2} The gluon, decomposed as a $q\bar{q}$ pair, is pulled
far from the original minimal surface of Fig. \ref{qqgloop}.}
\end{figure}

If the gluon line fluctuates to a position  not on the original minimal
surface, there are forces acting to pull it back.  Just as for the adjoint
Wilson loop of Fig. \ref{gluonloop}, these are not confining potentials (area
laws), but are due to perimeter and other forces, including color-Coulomb
forces.  Consider first a gross deformation of the Wilson loop of Fig.
\ref{qqgloop} (right), as shown in Fig. \ref{qqloop2}.  Although the areas
enclosed by contours 1 and 2 have increased, just as for Fig. \ref{qqgloop}
they are oppositely-oriented.  The only area law refers to the original 1-2
loop.  In fact, if one tries to pull the gluon a long way from the $q\bar{q}$
loop, the adjoint string discussed above will break, forming a gluon pair, and
the result will be a hybrid plus a glueball.

Evidently the techniques described here are applicable to the problem of
fluctuations of the area in a confining area law, although this fluctuation
problem is more complicated because one must treat an arbitrary number of
gluons, not part of the condensate of center vortices, exchanged between parts
of the Wilson line.  Originally, area-law surface fluctuations were thought of
as taking place in chromoelectric flux tubes \cite{lush}.  However, the center
vortex description of confinement is different, and requires a new approach. 
We discuss this a little more in the concluding section.

\subsection{Toward quantitative gluonic potentials}

If the two Wilson loops 1$G$ and $G2$ shown in the right side of Fig.
\ref{qqgloop} were statistically independent with respect to the center
vortices linking them, the expectation value in Eq. (\ref{hsplit} for the
hybrid propagator would factor, and we would have for this propagator
\begin{eqnarray}
\label{loopprod}
\Delta_H(x;y) & = & {\mathcal{N}}\int_0^{\infty}\int_x^y(dz)\exp
\{\frac{-M}{2}\int_0^s d\tau [\dot{z}^2+1]\}\langle TrU(1G)\rangle_A \langle
TrU(G2)\rangle_A \\ \nonumber
 & = & {\mathcal{N}}\int_0^{\infty}\int_x^y(dz)\exp
\{\frac{-M}{2}\int_0^s d\tau [\dot{z}^2+1]\}\exp [-K_F(A(1G)+A(G2)]
\end{eqnarray} 
where $\langle \cdot \rangle_A$ is the expectation value over center vortex
configurations and we have explicitly exhibited the gluonic path integral.  In
this equation $A(1G)$ is the minimal surface spanning contour 1$G$.  Actually,
the center vortex linkages in the two loops are not quite independent, because
the loops share the leg $G$, but the correlation effects are short-ranged (see
\cite{corn04} for a discussion) and will be ignored here.

Take the original Wilson loop 12 of Fig. \ref{qqgloop} to be a rectangle of
length $T$ in the Euclidean time direction $t$, and of length $R$ in the $z$
direction, centered at $z=0$. 
We will extract the non-relativistic potential from the path integral of Eq.
(\ref{loopprod}) as the coefficient of the length $T$ of the time-directed
segments of the original quarks paths 1,2, in the limit $\rightarrow \infty$. 
As mentioned previously, all potentials for hybrids will be expressed relative
to the potential in the original $q\bar{q}$ system, so we subtract $RT$ from
the area $A(1G)+A(G2)$ in determining the gluon potential.  If the gluonic
contour segment $G$ lies entirely in the original (flat) $q\bar{q}$ plane,
then the spanning
surfaces remain flat and the sum of areas $A(1G)+A(G2)$ is $RT$, so the
potential is zero.  But for small
transverse excursions of $G$ into the $x,y$ directions, the sum of areas is
slightly increased.   The simplest formula for this small change in the
minimal surface area involves presenting the coordinates perpendicular to the
surface, which we take to be $x,y$, as harmonic functions of two surface
variables, which we take to be $z,t$:
\begin{equation}
\label{area}
A(1G)+A(G2)-RT\simeq \frac{1}{2}\int_{\Gamma (1G)+\Gamma (G2)}
dtdz[x'^2+\dot{x}^2+y'^2+\dot{y}^2]
\end{equation}
where  the dot indicates a $t$
derivative, the prime indicates a $z$ derivative, and the integral is over the
flat surfaces $\Gamma (1G),\;\Gamma (G2)$ spanned by the original quark lines
and the projection of the gluon contour $G$ onto $x,y=0$.    At first sight
this is confusing, because the gluonic path variables are originally presented
as   a Euclidean four-vector
$z_{\mu}(\tau )=(x,y,z,t)$, with all four coordinates depending on the
proper-time varible $\tau$.  Because the gluon is assumed
non-relativistic, Euclidean time $t$ is essentially the same as proper time
$\tau$.  And the functions $x(z,t),y(z,t)$ required in the area integral of
Eq. (\ref{area}) are harmonic and therefore, in principle at least, determined
by their boundary
value on the contour $z_{\mu}(\tau )$. 

Note that the two transverse variables $x,y$ are just the variables used in
\cite{lush} to describe fluctuations of a surface formed from a chromoelectric
flux tube extended in time.  There should be, then, a close relation between
the dynamics of a valence gluon in a hybrid and surface fluctuations.  We will
not explore that relationship further here, but it is a very interesting
subject.

Next, we assume that the actual gluon paths $z_{\mu}(\tau )$ stay close to the
original minimal surface and are not highly-convoluted.  In that case, the
perturbed surface is nearly flat.  An example is a portion of the helicoid, a
minimal surface generated by rotating a rigid straight line segment around a
fixed axis while at the same time moving it along the axis.  Let the axis be
the segment of the quark line 1 in the $t$ direction, and the line segment at
$t=0$ is the $z$ axis from the origin to line 1.  This line segment has length
$R/2$.  The line segment when $t$ has a value scaling with the overall time
$T$ is slightly rotated around line 1 so that it passes through the point
$x=0,z=0,y$ with $y\ll R/2$.  The slightly-curved contour $z_{\mu}(\tau )$
is implicit in this description, and we need not be explicit; it is enough to
observe that only small errors are made by approximating all portions of the
contours by straight lines.  Because the contour goes from $x,y=0$ at $t=0$ to
small values $x,y$ at $t\sim T$, it is clear that $\dot{x} \sim
x/T,\;\dot{y}\sim y/T$.  Since $T\rightarrow \infty$ we can drop the time
derivatives.  By a similar line of argument, $x'\sim x/R,\;y'\sim y/R$. Then
the area integral in Eq.(\ref{area}) is of the form
\begin{equation}
\label{newarea}
A(1G)+A(G2)-RT\sim T[\frac{1}{2R}(x^2+y^2)].
\end{equation}
 We need not work out the full details, because it should be clear by now that
if we
approximate pieces of the full contour by straight lines we are simply
describing  physically the gluon as a bead that runs without friction along a
string stretched from $z=R/2$ to $z=-R/2$.   Transverse string fluctuations
stretch the string and give rise to the potential of Eq. (\ref{qqpot}) below. 
As is evident from Eq. (\ref{newarea}), the small-fluctuation potential is
that of a harmonic oscillator.

Before discussing this potential (and its $qqqG$ analog) we return briefly to
the perimeter terms that keep the gluon from going too far beyond the original
contour 12.  
These perimeter terms in the adjoint Wilson loop are discussed
in the center-vortex picture in \cite{fbo},\cite{corn98}.  These terms give
rise to a breakable string:  A potential which is approximately
linearly-rising, but goes flat when enough energy is stored in the string to
pop a gluon pair out of the vacuum.   It has been argued (see \cite{fbo} and
references therein) that $K_A$ is related to $K_F$ by Casimir scaling; another
elementary argument is that $K_A=2K_F$ because an adjoint string is two
fundamental strings.  The rough numerical estimates of \cite{corn98} yield a
$K_A$ that is bigger than $K_F$, but not quite according to either of these
arguments. In the present work it is good enough, within the
context of our approximations, to take $K_A=2K_F$.  

Note that the adjoint string necessarily breaks when an energy comparable to
2$M$ is stored in it.  This means that the more relativistic the gluon, the
closer one approaches to string breaking and mixing of the hybrid with states
containing another gluon.  We will not take this issue up here.

\section{Gluonic potentials}
\label{gluonpot}

Rather than continue to deal directly with fluctuations of a minimal surface,
as in Eq. (\ref{newarea}), we will find a very good approximation to the
potentials we need by starting from linearly-rising potentials attaching the
gluon to the quark lines.  Of course, these potentials cannot be taken
literally when the gluon fluctuates far from the original minimal surface, and
only their expansions for small fluctuations are of interest.  As we saw in
the last section, these give rise to harmonic oscillators with frequency $\sim
[K_F/(MR)]^{1/2}$.  Small fluctuations mean small gluon velocities, and so
this regime of small fluctuations should be non-relativistic, if the gluon
mass is large enough.

Fot the $q\bar{q}G$ hybrid the gluon can move freely along the string, except
that it  cannot pass beyond the
string endpoints.    The free longitudinal momentum is associated with a
momentum $p$ in this surface of the form $p=N\pi/R$.   If $N>1$ the motion is
relativistic for $R$ is less than about 2 fm, and so we will concern ourselves
only with the ground state $N=1$. Details of an approximate non-relativistic
analysis of the combined longitudinal and transverse Schr\"odinger equation
are in the Appendix.

  As we will see, the numbers turn out such that even for the gluonic ground
state of a hybrid relativistic corrections seem to be important, and we must
ask where and how relativistic effects enter.   We find in the Appendix that
the gluon velocity $\beta$ obeys  $\beta^2 \sim [K_F/(M^3R)]^{1/2}$, and 
relataivistic effects are important for $R\simeq$ 0.6 fm.

Although we do not study mesonic hybrids in any detail in this paper, it is
worth giving the general scheme of the gluonic potential for the $q\bar{q}G$
system, both for future use and because the same themes come up for the $qqqG$
potential.

\subsection{Mesonic hybrid potential}

\begin{figure}
\includegraphics[height=3in]{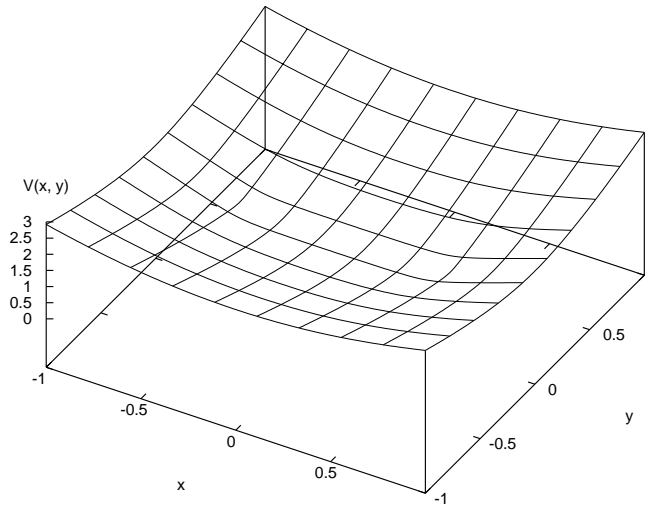}
\caption{\label{pot1} The potential of Eq. (\ref{qqpot}), with $R$ and $K_F$ 
scaled to unity.  The $y$=0 line is flat from $x=-1/2$ to $x=1/2$.}
\end{figure}

For the $q\bar{q}$ system the arguments of the previous section lead to a 
non-relativistic potential that is a
standard linearly-rising potential for each of the areas 1$G$ and $G$2 in Fig.
\ref{qqgloop}:
\begin{equation}
\label{qqpot}
V(\vec{R}, \vec{r}) =
K_F\{[(z-R/2)^2+\rho^2]^{1/2}+[(z+R/2)^2+\rho^2]^{1/2}-R\}
\end{equation}
[This potential has been proposed before \cite{leya} on purely phenomenological
grounds.]
This potential has all the properties expected from our discussion in the
previous section.  In the minimal surface of the $q\bar{q}$ system, which is
$\rho=0,\;|z|<R/2$, the potential is identically zero.  For $|z|>R/2$ there 
are linearly-rising terms beginning at $|z|=R/2$ and extending outside the
minimal surface, as explained in the Appendix.  We identify these terms with
the breakable adjoint string, although the potential of Eq. (\ref{qqpot}) does
not explicitly show breaking.  For small excursions in $\rho$ (transverse to
the minimal surface) we expand the potential in $\rho$.  The linear terms
vanish, as they must (except at the endpoints $|z|=R/2$), and near $z\simeq 0$
the quadratic
terms are of the form $(2K_F\rho^2/R$, with an oscillator frequency $\Omega
(R)\simeq 2[K_F/(MR)]^{1/2}$, where $R$ is the $q\bar{q}$ separation. As $z$
approaches the endpoints $\pm R/2$, the energy needed for a given transverse
displacement grows, so there is a force keeping the gluon ``bead" away from
the endpoints unless the string on which the bead rides is not stretched at
all.  In Fig. \ref{pot1} we plot the
potential for the $d=3$ case, with $q$ and $\bar{q}$ separated by unit distance
along the $x$ axis and $y$ as the transverse coordinate. 

For further discussion of this potential, see the Appendix.

\subsection{Baryonic hybrid potential}

In general, the potential is the string tension $K_F$ times the length of some
minimal string network joining the original three quarks to the $q\bar{q}$
pair of the gluon.
We will see that in principle this potential requires consideration of minimal
string lengths with different functional forms in different regimes.  Once the
work of expanding the potential in various regimes is all done, we get a
result which differs only slightly from the result gotten from the
intuitively-appealing potential
\begin{equation}
\label{newqqqpot1}
V(\vec{R}_1,\vec{R}_2,\vec{R}_3;\vec{r})=K_F\{|\vec{R}_1-\vec{r}|+
|\vec{R}_2-\vec{r}| +|\vec{R}_3-\vec{r}|\}.
\end{equation}
The reader who is willing to believe this can go straight to the end of this
section, where the final results for the harmonic terms are given.  We will
only work out the case when the three quarks (at fixed time) are at the
corners of an equilateral triangle.  This makes little difference in comparing
theory with data, since \cite{tsuga} the corresponding lattice data are not
much different for different quark triangles, provided that they have the same
length $L_{min}$ of the Y-shaped Steiner string network joining them.

\begin{figure}
\includegraphics[height=3in]{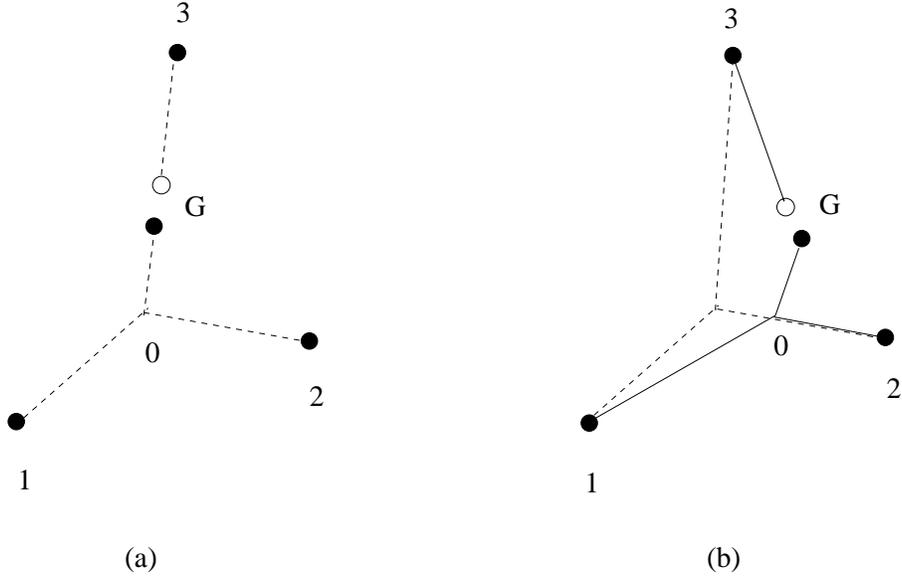}
\caption{\label{qqqtop} A view of the $qqqG$ system looking along the quark
world lines. The original  Steiner network for no gluon is shown as dotted
lines meeting at 0. (a)  The gluon lies along the line 03; there is no
resulting potential (except at the end points).  (b)  The gluon lies
in regime 1 (defined in the text); the original Y strings are deformed into a
longer Steiner network,
raising the potential from zero.}
\end{figure}

Fig.  \ref{qqqtop} shows a view of a $qqqG$ system looking down along the $t$
axis on the
transverse plane, taken to be the $xy$ plane.  The original quark system 123
has no angle as great as
$2\pi
/3$ and so has an internal Steiner point; this is the only type of
configuration we will consider here.  Full circles represent the quarks 1, 2,
3 of Fig. \ref{qqqloop} and the quark in the split gluon line, and the
unfilled circle is the antiquark of that line.  Dotted lines represent the
original Y-shaped strings joining the quarks in the absence of the gluon;
generically they
meet at the point 0 in Fig. \ref{qqqtop}(a) at  an angle of $2\pi /3$. 
Splitting the gluon leads to a new baryon 12$G$, which is joined by its own
string network.  

We have chosen to single out quark 3 for special treatment in forming the
baryonic hybrid.  This could be appropriate if the quarks had different
flavors or masses.   Later, we will average the potential over interchanges of
the three quarks to yield an energy symmetric in the quark labels. This is
equivalent to dropping certain $l=2$ terms in the potential.

The essential point to keep in mind for the following discussion is that if a
triangle has no angle as great as $2\pi /3$, it has a Steiner point inside the
triangle where the three strings meet at an angle of $2\pi /3$.  But if an
angle of the triangle is $2\pi /3$ or greater, the minimal-length string lies
along the two sides meeting at that angle. Because of this, the algebraic form
of the potential changes, depending on
which of several regimes the gluon occupies.  Fluctuations of the gluon in the
out-of-plane ($z$) direction decouple from fluctuations in the $XY$ plane, and
are easily treated separately.  If the gluon is fluctuating in-plane, we need
to consider two regimes.   Regime 1 is the sum of triangles 103 and 203 in the
figure; regime 2 is the triangle 012.

Consider first regime 1.  For the special position shown
in  Fig. \ref{qqqtop}(a), with the gluon sitting on the original Steiner line
03, the strings for the new
baryon coincide with those of the original.  This is because the angle 102  is
precisely $2\pi /3$.  This, plus our remarks already made for a gluon lying in
a mesonic surface, show that  there is no potential energy associated with
moving the gluon along the line 03, except within a QCD length of the ends 0
and 3.  Let us take the line 03 to lie in the $y$ direction; then we expect to
(and do) find only oscillations in the $x$ direction.

The potential to be expanded for all of regime 1 (again referenced to the
original $qqq$ potential) is
\begin{equation}
\label{newqqqpot}
V(\vec{R}_1,\vec{R}_2,\vec{R}_3;\vec{r})=K_F\{L_{min}(R_{1G},R_{2G},R_{12})
+|\vec{R}_3-\vec{r}|-L_{min}(R_{12},R_{13},R_{23})\}
\end{equation}
where $R_i$ is the vector position of the $i^{th}$ quark, $\vec{\rho}$ is the
in-plane
position of the gluon, $R_{ij}$ is the distance between quarks $i$ and $j$,
and $R_{iG}$ the distance between quark $i$ and the gluon.  Also,
$L_{min}(a,b,c)$ is the length of the Steiner network joining three points that
form a triangle of sides $a,b,c$.  The expression for this minimal length is:
\begin{equation}
\label{steinerlength}
L_{min}(a,b,c)=\frac{1}{\sqrt{2}}\{(a^2+b^2+c^2)+
\sqrt{3}[4a^2b^2-(a^2+b^2-c^2)^2]^{1/2}\}^{1/2}.
\end{equation}

We take $\vec{\rho}=(x,0)$ and find the quadratic term in the potential
\begin{equation}
\label{xpot1}
\frac{9}{4}\frac{K_Fx^2}{L_{min}}
\end{equation}
where $L_{min}$ is the minimal length of the Steiner network joining the
original three quarks.  [If we had expanded the intuitive potential of Eq.
(\ref{newqqqpot1}) we would have found an equal contribution for $y$
oscillations.]  We will replace this by an average over choosing any of the
quarks as having the gluon antiquark joined to it, instead of quark 3 as in
the figure.  This just amounts to isotropizing the result of Eq.
(\ref{xpot1}) in the $x--y$ plane, with a resulting contribution from regime 1
to  the potential
of
\begin{equation}
\label{reg1}
V_1=\frac{9}{8}\frac{K_F}{L_{min}}(x^2+y^2).
\end{equation} 
This isotropization drops some $l=2,\;m=2$ terms in the potential. 

Regime 2 is the triangle 012 of Fig. \ref{qqqtop}(a).  If the gluon fluctuates
(in-plane) into this triangle, the baryon triangle 1$G$2 has an angle greater
than $2\pi /3$.  As a result, the minimal network is the sum of strings
running from the original quarks to the gluon, and the potential is indeed
that of the simple intuitive form in Eq. (\ref{newqqqpot1}). This potential
has the property that it is  symmetric with respect to
interchange of quark labels.   It has an equilibrium point at the origin,
reflecting the nature of the origin as a Steiner point.

We give the expansion of the regime-2 potential for generic interquark
distances, for future studies.   
Choose the (two-dimensional) quark vectors as
\begin{equation}
\label{quarkv}
\vec{R}_1=R_1(\frac{-\sqrt{3}}{2},\frac{-1}{2});\;\vec{R}_2=
R_2(\frac{\sqrt{3}}{2}, \frac{-1}{2});\;\vec{R}_3=R_3(0,1).
\end{equation} 
With this coordinatization the Steiner point 0 is at the origin.  The expansion
of the potential of Eq. (\ref{newqqqpot}) in transverse coordinates is
\begin{equation}
\label{potexp}
V_2(\vec{R}_1,\vec{R}_2,\vec{R}_3;\vec{\rho})=\frac{K_F}{2}\{\rho^2\sum
\frac{1}{R_i}
-\sum \frac{[\hat{R}_i\cdot \vec{\rho}]^2}{R_i}\}+\cdots
\end{equation}
where $\vec{\rho}=(x,y)$. 
Working out the algebra, using Eq. (\ref{quarkv}), yields
\begin{equation}
\label{potexp1}
V_2(\vec{R}_1,\vec{R}_2,\vec{R}_3;\vec{r})=\frac{K_F}{2}\rho_iV_{ij}\rho_j
\end{equation}
where the matrix $V_{ij}$ has eigenvalues $\lambda_{\alpha}$:
\begin{equation}
\label{eigenv}
\lambda_1,\;\lambda_2= \frac{1}{2}\{\sum
\frac{1}{R_i}\pm [\sum \frac{1}{R_i^2}-\sum_{i\neq
j}\frac{1}{R_iR_j}]^{1/2}\}.
\end{equation}
This energy $\varepsilon (R_1,R_2,R_3)$  is clearly not a function of the
minimal distance
$L_{min}=R_1+R_2+R_3$.  The deviations from dependence only on $L_{min}$ are
acceptably small as long as none of the $R_i$ are too small, which in any case
is necessary for all of our developments.    For the equilateral triangle, the
regime-2 potential is
\begin{equation}
\label{reg2}
V_2=\frac{9}{4}\frac{K_F}{L_{min}}(x^2+y^2).
\end{equation}

Finally, the out-of-plane oscillations come from expanding Eq.
(\ref{newqqqpot1}) for a gluonic excursion in $z$.  The result is
\begin{equation}
\label{regz}
V_z=\frac{9}{2}\frac{K_F}{L_{min}}z^2.
\end{equation}

To construct the final potential, we note that regime 1 is twice the size of
regime 2.  For a simple  harmonic-oscillator wave function as a variational
trial wave function this suggests that we take the in-plane potential as
\begin{equation}
\label{vinplane}
V_{in}=\frac{2}{3}V_1+\frac{1}{3}V_2 =
\frac{3}{2}\frac{K_F}{L_{min}}(x^2+y^2).
\end{equation}
The total potential is 
\begin{equation}
\label{fullpot}
V_{in}+V_z\equiv V(\vec{x})=\frac{\beta_zK_F}{L_{min}}z^2+
\frac{\beta_{\rho}K_F}{L_{min}}(x^2+y^2)
\end{equation}
where $\beta_z=9/2,\;\beta_{\rho}=3/2$.  This potential can be isotropized in
$z$ but it makes little numerical difference.

Strictly speaking, the harmonic potentials in $V$ terminate at the positions of
the quarks in the original baryon, and are replaced by something more
complicated (essentially the breakable adjoint string).  However, in the
regime of large interquark separations this is not important.  The reason is
that if the unadorned harmonic potentials are used, the (non-relativistic)
range of the gluon wave function is of order $[L_{min}/(MK_F)]^{1/4}$.  This
is small compared to $L_{min}$ when $L_{min}\gg a$, with
$a=(2K_FM)^{-1/3}\simeq $ 0.33 f.  Even if this termination of harmonic
potentials is taken into effect, it is a small effect numerically.

The non-relativistic ground-state energy
of the total
potential $V$, plus the gluon mass $M$, yields a string energy for the baryon
hybrid of
\begin{equation}
\label{final}
\varepsilon_{string} (L_{min})=
M+(\frac{3}{2}+\sqrt{3})[\frac{K_F}{ML_{min}}]^{1/2}\equiv \xi
[\frac{K_F}{ML_{min}}]^{1/2};\;\xi \simeq 3.23.
\end{equation}
To this we add the color-Coulomb energy difference, estimated in the Appendix
as  
 \begin{equation}
\label{colorcoulomb}
V_c=\alpha_c\{\frac{\sqrt{3}}{L_{min}}-\frac{6}{\sqrt{L_{min}^2+9a^2}}\}
\end{equation} 
with $\alpha_c\simeq $ 0.15 \cite{tsuga} and $a$, defined above, is $\simeq$
0.33 f.  (The length $a$
accounts for spreading of the gluon wave function, as described in the
Appendix.)  

\section{Comparison of theory with lattice data}
\label{datatheory}

\begin{figure}
\includegraphics[height=3in]{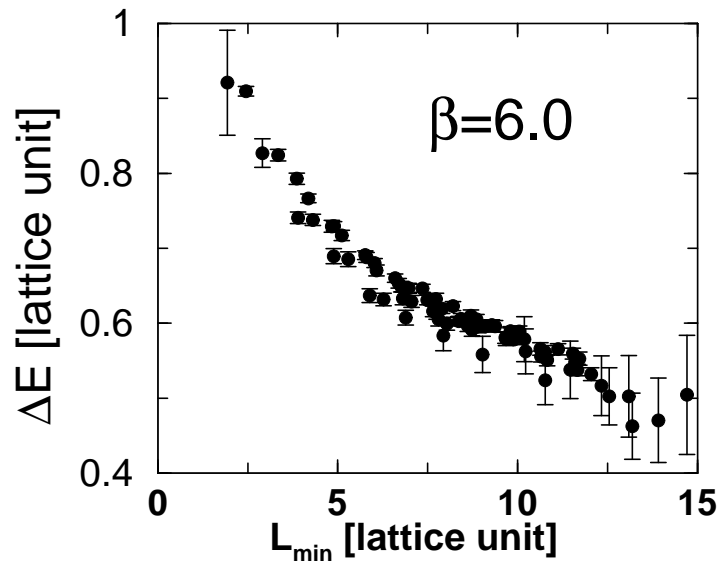}
\caption{\label{suga} Lattice determination of the single-gluon excitation
energy $\Delta E$ of a heavy $qqqG$ system, as a function of $L_{min}$, the
length of the Steiner path joining the three quarks. From \cite{tsuga}.}
\end{figure}

There are extensive heavy-quark lattice data available on the heavy-quark
$qqqG$ system
\cite{tsuga}.  Fig. \ref{suga} shows the measured gluonic
excitation energy $\Delta E$ of the gluon in the $qqqG$ system, measured
relative to the $qqq$ system, and displayed as a function of the length of
strings joining the original three quarks \cite{tsuga}.  Although in principle
the energy depends on three separate
interquark distances, \cite{tsuga} finds that their lattice results fall
pretty much on a single curve, parametrized by the minimum length $L_{min}$ of
the Y-string joining the original $qqq$ system.   Both the energy and
distance axes are scaled in lattice units, which for \cite{tsuga} is about 0.1
fm.  The energies in GeV are approximately twice the numerical values shown on
Fig. \ref{suga} and on the next figure. 

\begin{figure}
\includegraphics[height=5in]{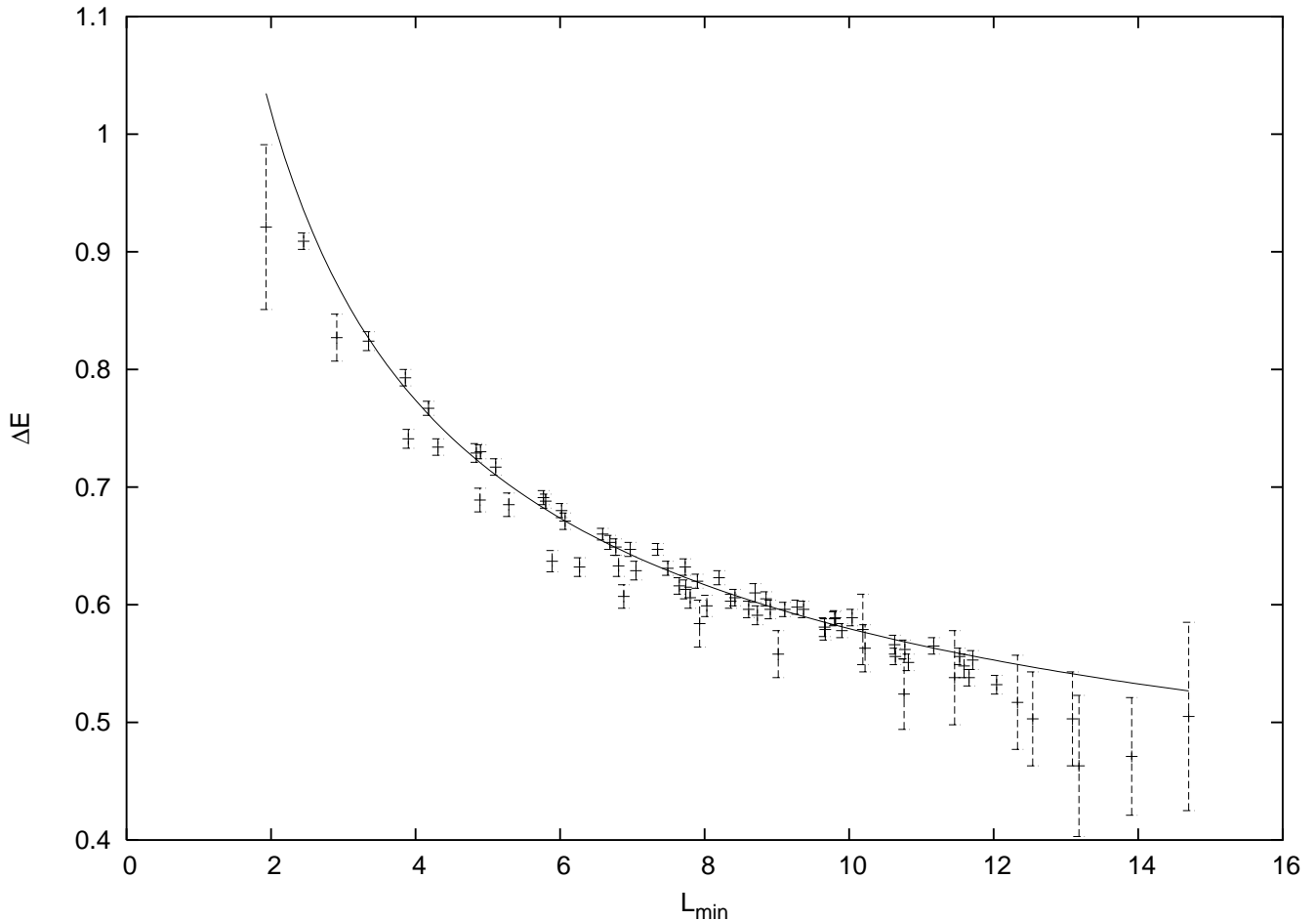}
\caption{\label{theorydata} Data (again in lattice units) of Fig. \ref{suga}
shown with the theoretical
calculation discussed in the text.}
\end{figure}

We can match the data of Fig. \ref{suga} reasonably well, using the variational
treatment of the anisotropic harmonic-oscillator Schr\"odinger equation with
relativistic kinetic energy, given in the Appendix, and a gluon mass
$M$ of around  500 MeV.  Coulomb effects are added perturbatively, as described
in the Appendix. Since the data themselves fall roughly on a single curve for
various $qqq$ geometries with the same $L_{min}$, we have done numerical
calculations only for the special case of an equilateral triangle.  These
calculations and the lattice data \cite{tsuga} are shown in Fig.
\ref{theorydata} for the value $M$ =  500 MeV.  

The fit is surprisingly good at $L_{min}\geq$ 4, which is a distance short
enough for relativistic effects to be important.  In fact, the total energy at
this distance is about $3M$, roughly apportioned equally among mass energy,
potential energy, and kinetic energy.  For smaller $L_{min}$ there really
should be a fair amount of
mixing with states containing more than one gluon, such as a baryon plus a
glueball.  We do not know to what extent such mixing is or is not accounted
for in the lattice computations.  The mass value of  500 MeV is a little small
compared to a typical lattice value \cite{deforc} of 600 MeV, but certainly
falls within an acceptable range of determinations of $M$.  A proper handling
of mixing of higher angular momentum states might raise the fitted gluon mass
somewhat.

\section{Conclusions}

We showed that there is a straightforward path from the fundamental elements of
center-vortex QCD to a description of the  gluonic ground state of heavy-quark
$qqqG$ hybrids (and, for that matter, of $q\bar{q}$ hybrids, although we have
not attempted any comparison with lattice data for these).  In graphic terms,
the result is that the valence gluon can be thought of as a massive bead more
or less freely running along a string joining the quarks, and subject to
transverse restoring forces as the string is stretched by gluonic
fluctuations.  

The hybrid problem simplifies at large interquark separation $R$, which is the
regime we treated explicitly.  Although our treatment, based on approximations
to a Schr\"odinger equation with  fully-relativistic kinetic energy is not
exact, it is not subject to the usual burden of purely
phenomenological assumptions made about the nature of confining potentials and
so forth. The fundamental elements used were: 1) Splitting of a gluon
proper-time adjoint Wilson line into a pair of co-located and
oppositely-directed fundamental Wilson lines, which join to  the three other
Wilson lines in a specific way.  2)  Recognition that for a massive gluon
the equal-time Schr\"odinger equation applies for sufficiently large $R$, but
that relativistic corrections to the kinetic energy are important for
physically-interesting quark separations.

We used the gluon mass $M$ as a single parameter for fitting the lattice data
of Ref. \cite{tsuga}, and found a pretty good fit for a mass of about  500 MeV. 
This mass is a little less than the usual value of about 600 MeV found on the
lattice, but at this stage of the investigation seems quite satisfactory.  

One could, in principle, think of extending  this work to excited gluonic
states in hybrids.  However, at any reasonable interquark separation, the
excited states are relativistic, and there will be some mixing with states
with more than one gluon.  As discussed in the Appendix for $q\bar{q}G$
hybrids, if this mixing is ignored there should be several contributions to
the gluonic energy, including a stringlike energy of the form $N\pi/R$, as
invoked by \cite{jkm,juge} for such hybrids, plus terms of order
$(K_F/R)^{1/3}$ and Coulomb terms.   

We noted that the treatment of the valence gluon by adjoint-line splitting
suggested a possible path toward describing the fluctuations of surfaces
spanning fundamental Wilson loops.  Originally, these surfaces were envisaged
\cite{lush} as having an essentially tangible existence as chromoelectric flux
tubes.  But fluctuations of spanning surfaces in the area law is more
difficult to think about in the center vortex picture, where   the description
of confinement begins with the topological statement that the area law arises
from the fluctuations in phase factors (elements of the center group)
associated with topological linking of center vortices with the Wilson loop. 
At this level of description of confinement, the area in an area law is a
minimal surface spanning the Wilson loop and is not subject to fluctuations
\cite{corn04}.  However, coupling of gluons that are not part of the
condensate of confining vortices can give rise to surface fluctuations, as we
have seen in the present paper for valence gluons in hybrids.  Work is
underway to understand the description of such fluctuations when a large
number of such gluons are coupled to a Wilson loop.

Finally, we have not discussed here the mixing of the valence gluon in a hybrid
with genuine $q\bar{q}$ pairs (as opposed to the fictitious pairs produced by
splitting an adjoint Wilson line).  If the valence gluon in $qqqG$ is replaced
by a valence $q\bar{q}$ pair, the result is a special kind of pentaquark, or
$qqqq\bar{q}$ system. This pentaquark is definitely not the same as the famous
pentaquark \cite{zhu}, which is of the form $uudd\bar{s}$, where the antiquark
cannot pair with any quark to form a flavor-neutral state such as a gluon. 
Nonetheless, the problem of baryonic hybrid mixing with pentaquarks is of some
interest, and worthy of future study.

\begin{acknowledgments}
This work was supported in part by  NASA ATP grant  NAG5-13399.
\end{acknowledgments}

\appendix

\section{Treatment of the Schr\"odinger equation}

Although the gluon has a mass $M$, it is not heavy enough to insure
that the non-relativistic  Schr\"odinger equation applies throughout the range
of quark separations for which lattice data are availaable.  We give first a
simple variational treatment of the ground-state excitation energy associated
with the harmonic potentials encountered for hybrids, including relativistic
effects for the kinetic energy.  Then we go on to sketch a treatment of the
gluon as a bead sliding freely along a string, and confined harmonically in
the transverse direction.  This is the problem encountered for the $q\bar{q}G$
hybrid and for one of the strings in the $qqqG$ hybrid.  Finally, we briefly
sketch a treatment for color-Coulomb corrections.

\subsection{Relativistic harmonic  oscillator for the baryonic hybrid}

The string potential for the $qqqG$ state is the sum of Eqs. (\ref{regz}, 
\ref{vinplane}), and has the form
\begin{equation}
\label{generalpot}
 V(\vec{x})= \frac{\beta_zK_F}{L_{min}}z^2+
\frac{\beta_{\rho}K_F}{L_{min}}(x^2+y^2)
\end{equation}
where $\beta_z=9/2,\;\beta_{\rho}=3/2$   and $L_{min}$ is  the minimum length
of the Steiner network.  The full Hamiltonian, including relativistic kinetic
energy, is
\begin{equation}
\label{fullh}
H=\sqrt{\vec{p}^2+M^2}+V(\vec{x}).
\end{equation}
We take a simple normalized variational wavefunction $\psi$:
\begin{equation}
\label{varpsi}
\psi
(\vec{x})=(\frac{\alpha_z}{\pi})^{1/4}(\frac{\alpha_{\rho}}{\pi})^{1/2}\exp
\{-\frac{1}{2}[\alpha_{\rho}( x^2+y^2)+\alpha_zz^2]\}
\end{equation}
and calculate the expectation value $\langle H \rangle \equiv \varepsilon
(\alpha_z, \alpha_{\rho} )$:
\begin{equation}
\label{hexp}
\langle H \rangle =
\frac{2}{\alpha_{\rho}\sqrt{\alpha_z\pi}}\int_0^{\infty}dp_{\bot}p_{\bot}
\int_0^{\infty}
dp_{\|}(p_{\bot}^2+p_{\|}^2+M^2)^{1/2}\exp
\{-\frac{p_{\|}^2}{\alpha_z}-\frac{p_{\bot}^2}{\alpha_{\rho}}\}
+\frac{
K_F}{2L_{min}}(\frac{\beta_z}{\alpha_z}+\frac{2\beta_{\rho}}{\alpha_{\rho}}).
\end{equation}

The variational parameters $\alpha_{\rho ,z}$ are to be determined by
minimizing
$\varepsilon$, which can only be done numerically. 
Of course, the non-relativistic limit is easy to do, yielding
\begin{equation}
\label{limits}
\langle H\rangle  =  M+(\frac{ K_F}{2ML_{min}})^{1/2}
[\sqrt{\beta_z}+2\sqrt{\beta_{\rho}}].
\end{equation}

In the relativistic regime, one finds
\begin{equation}
\label{rellimit}
\varepsilon = \frac{3}{2}(\frac{3\beta K_F}{2\pi L_{\min}})^{1/3}.
\end{equation}
where $\beta_{rel}$ is a weighted average of $\beta_z$ and $\beta_{\rho}$,
approximately given by $\beta_{rel} =(1/3)(\beta_z+2\beta_{\rho})$.
The separation between the relativistic and non-relativistic regimes  is at a
critical interquark distance $R_c$, where
\begin{equation}
\label{rsubc}
R_c \sim \frac{\beta K_F}{M^3}.
\end{equation}
 When $R\gg R_c$ the problem is
non-relativistic.  Typically $R_c$ is perhaps 0.8 f. 

The fit to data of Fig. \ref{theorydata} comes from a combination of numerical
and analytical work on minimizing expressions such as in Eq. (\ref{hexp}).  It
turns out to be accurate enough to isotropize the potential of Eq.
(\ref{generalpot}, setting $x^2,\;y^2$ and $z^2$ all to $r^2/3$, and also to
take $\alpha_z=\alpha_{\rho}\equiv \alpha$.  Then a good analytic fit can be
made to the kinetic energy term, and the result is 
\begin{equation}
\label{epsapprox}
\varepsilon (\alpha )\simeq [\gamma \alpha + M^2]^{1/2}+\frac{3\beta 
K_F}{2\alpha R} 
\end{equation} 
with
\begin{equation}
\label{gambeta}
\gamma \simeq 1.38,\;\beta \simeq 2.52.
\end{equation}
The values of $\gamma$ and $\beta$ are correlated so as to give the exact
non-relativistic energy at large $R$.
It is now straightforward if lengthy to find an analytic expression for the
minimum over $\alpha$ of $\varepsilon (\alpha )$, involving the root of a
quartic equation.

\subsection{Line of equilibrium}

  We continue with a treatment of the Schr\"odinger equation that applies when
there is a line of equilibrium, not just a point.

To be specific, consider a
a $q\bar{q}$ state,  where the potential has the form in Eq. (\ref{qqpot})
(repeated for convenience)
\begin{equation}
\label{appglupot}
V(\vec{R}, \vec{r}) =
K_F\{[(z-R/2)^2+\rho^2]^{1/2}+[(z+R/2)^2+\rho^2]^{1/2}-R\}
\end{equation}.
Here (see Fig. 1) $\vec{R}=(0,0,R)$ is the separation vector of the heavy quark
and
antiquark, $\vec{r}= (x,y,z)$ is the gluon position
vector, and $\rho^2=x^2+y^2$ is the (square of the) gluon excursion out of the
minimal surface.  As discussed earlier, this is not literally the true
potential, but
it is accurate enough for our purposes.  

There is a second scale length $a$ in the problem, given by 
\begin{equation}
\label{aeqn}
a = (2K_FM)^{-1/3}\simeq 0.33 fm,
\end{equation}
 and we will require $R\gg a$.  Since $R_c/a\sim K_F/M^2<1$,  this requirement
puts us in the non-relativistic regime, and means that our considerations may
hold for the ground state, but not necessarily for excited states.

  We will see that the scale length for gluon excursions away from the minimal
$q\bar{q}$ surface is at most of order $(Ra^3)^{1/4}\ll R$.  Therefore we
provisionally take $\rho^2/R^2$ as an expansion parameter  for the potential. 
Such an expansion requires $|z|\ll R/2$, which we assume for now.  The first
two terms of the expansion are
\begin{equation}
\label{largerpot}
V=\frac{2K_F\rho^2}{R}  + \frac{8K_Fz^2\rho^2}{R^3}+\dots
\end{equation} 
There are no terms in the expansion of the form $z^N$; all powers of $z$ are
accompanied by powers of $\rho^2$.    Note that if $\rho \neq 0$, the point
$z=0$ is the equilibrium point,
midway between the quark and antiquark.  

 The expansion above fails for $|z|>R/2$.  To characterize this regime we set
$\rho=0$ in the potential, and find that the potential identically vanishes
for $|z|<R/2$, while for $|z|>R/2$
there are linearly-rising potentials which confine the $z$ motion.  
We will adopt the following essentially variational strategy, accurate enough
for present purposes:
Assume the ground-state wave function $\psi (\vec{r})$ for the Schr\"odinger
equation is separable, of the form
\begin{equation}
\label{sep}
\psi = \phi(z)(\frac{\alpha}{\pi})^{1/2}\exp [-\frac{\alpha \rho^2}{2}]
\end{equation}
where $\phi (z)$ is a normalized solution to the Schr\"odinger equation at
$\rho =0$.  The other factor is a harmonic-oscillator ground state, and we
will determine the parameter $\alpha $ from the potential of Eq.
(\ref{appglupot}), averaged over the $z$ motion. That is, the effective
potential for $\rho $ is given by
\begin{equation}
\label{rhopot}
V_{eff}(\rho )=\int_{-R/2}^{R/2}dz|\phi (z)|^2V(\rho ,z).
\end{equation}

 We now construct a good approximation to $\phi (z)$. For
$z>0$ the potential is accurately given by
\begin{equation}
\label{linris}
V(z)\simeq 2K_F|z- (R/2)|\theta (z-(R/2)),
\end{equation}
and the region $z<0$ contributes symmetrically. The coefficient 2$K_F$ is a
decent approximation to  the true adjoint string tension.  Of course, the
adjoint string is breakable, and the potential of Eq. (\ref{linris}) is
applicable only for sufficiently small $|z-(R/2)|$.  The ground-state solution
to
the
Schr\"odinger equation which is symmetric under $z\rightarrow -z$ is 
\begin{equation}
\label{linsoln}
\phi (z) = N_1 \cos (qz)\theta [(R/2) - |z|] + \{N_2 Ai (\xi )\theta [z -
(R/2)]
+ z\leftrightarrow -z\}.
\end{equation} 
Here $N_{1,2}$ are normalization constants, $q$ is the momentum, and $Ai(\xi )$
is an Airy function of argument 
\begin{equation}
\label{xidef}
\xi = \frac{1}{a}(z-\frac{R}{2})-q^2a^2.
\end{equation}
The $z$-motion 
energy eigenvalue is $\varepsilon_z=q^2/(2M)$, with the momentum $q$ determined
by the
matching condition
\begin{equation}
\label{match}
-qa \tan (\frac{qR}{2})=\frac{Ai'(\xi )}{Ai(\xi )}|_{\xi = -(qa)^2}.
\end{equation}
At $R=0$ this yields the well-known result $(qa)^2\simeq \xi_0$  where $-\xi_0
=-1.02+$ is the first zero of $Ai'(\xi )$. 
For large $R$ one easily sees that $q$ must decrease toward $\pi /R$, so that
$\varepsilon_z\simeq \pi^2/(2MR^2)$.  Analysis which we do not give here
suggests that the
following form should be a good representation of the momentum $q(R)$ and
$\varepsilon_z(R)$ as  functions of $R$:
 \begin{equation}
\label{evsr}
q(R) \simeq \frac{\xi_0^{1/2}}{a+\gamma R};\;
\varepsilon_z(R)\simeq \frac{\xi_0}{2Ma^2(1+\gamma R/a)^2}
\end{equation}
where $\gamma = \xi_0^{1/2}/\pi$.  This correctly yields the large-$R$ behavior
and is  correct to about 20\% for small $R$.  

After some uninteresting algebra, one finds that at large $R$ the Airy
functions' renormalization constant $N_2$ is small of order $a/R$, and
therefore in calculating the $\rho$ potential from Eq. (\ref{rhopot}) we need
the integral
\begin{equation}
\label{rhopot2}
V_{eff}(\rho
)=K_F\int_{-R/2}^{R/2}dz\frac{2}{R}\cos^2(qz)\{[(z-R/2)^2+\rho^2]^{1/2}+
[(z+R/2)^2+\rho^2]^{1/2}-R\}.
\end{equation} 
We have calculated this integral numerically, and find that for small $(\rho
/R)^2$ it is reasonably well-approximated by the $z=0$ expansion already given
in Eq. (\ref{largerpot}):  $V_{eff} \simeq 2K_F\rho^2/R$.  The reason is that
in the integral the weight function $\cos^2(qz)$ is small near the end points. 
Now the parameter $\alpha$ of the transverse wave function in Eq. (\ref{sep})
is  determined ($\alpha = 2(K_FM/R)^{1/2}$), and the total non-relativistic
energy at large $R$ is 
\begin{equation}
\label{hybridenergy}
\varepsilon(R) \simeq M+2 (\frac{K_F}{MR})^{1/2} +\frac{2\xi_0}{2Ma^2(1+\gamma
R/a)^2}.
\end{equation}
For this oscillator one has $\langle \rho^2 \rangle \simeq (Ra^3)^{1/2}$ which,
as mentioned above, justifies expanding the full potential in powers of $(\rho
/R)^2$ at large $R$.  

Excited-state solutions to the gluon motion in the $z$ direction have momentum
$q$ given by $\pi N/R$.  When $N>MR/\pi$ the $z$ motion is relativistic, with
energy approximately equal to $q$.  The transverse motion may be relativistic
as well, requiring consideration of string breaking by gluon-pair formation. 
Without taking this phenomenon into account, one may expect corrections to the
string energy $N\pi /R$ to scale [as given in Eq. (\ref{rellimit})] with
$R^{-1/3}$.   However, many other effects come into play, and we have not
attempted to analyze data, such as that of \cite{juge}.

\subsection{Color-Coulomb corrections}

  We will
treat the Coulomb potential as a perturbation to the Schr\"odinger equation. 
If this potential were uniformly of  order $\alpha_c/R$ for typical interquark
separation $R$,  treating it as a
perturbation would be valid (non-relativistically) when  
\begin{equation}
\label{coulcond}
R \gg \frac{\alpha_c^2M}{K_F}.
\end{equation}
 which is easily satisfied for $R$ values of concern to us.

\subsubsection{Baryonic hybrid}

In the baryonic hybrid $qqqG$, the minimum of the gluon
potential is at the Steiner point, so it does not get close to the quarks, at
least when these are well-separated ($R\gg a$).  When the quarks get close, so
that $R\leq a$, the potential ceases to be $1/R$, due to smearing by the gluon
wave function.  The net color-Coulomb potential for the quarks, which can be
treated classically, in the original baryon is to be subtracted from the
Coulomb potential for the hybrid.  To a good enough approximation, the net
Coulomb potential for our standard equilateral triangle is
\begin{equation}
\label{colorcoul}
V_c=\alpha_c\{\frac{\sqrt{3}}{L_{min}}-\frac{6}{\sqrt{L_{min}^2+9a^2}}\}.
\end{equation} 
We will use $\alpha_c\simeq$ 0.15, appropriate for the lattice with $\beta$ =
6.0 
and quenched quarks.

\subsubsection{Mesonic hybrid}

In the mesonic hybrid the gluon can get close to a quark even when these are
well-separated, and the above
argument does not straightforwardly apply when the gluon position $z$ aliong
the string is near $\pm R/2$, where
the distance scale in the gluon-quark Coulomb potential is not generically of
order $R$. Nonetheless, we will find that the probability that the gluon
approaches the ends
of its string, where the Coulomb potential is not of order $\alpha_c/R$, 
scales with $1/R$ so that in the end the Coulomb contribution to the energy
does scale with $1/R$.

The Coulomb potential $V_C$ (relative to the $q\bar{q}$ potential with no
gluon) is
\begin{equation}
\label{coulomb}
V_C = \alpha_c \{\frac{1}{R}
-[(z-R/2)^2+\rho^2]^{-1/2}-[(z+R/2)^2+\rho^2]^{-1/2}\}.
\end{equation}
    
 A reasonable approximation is to replace the terms $z\pm (R/2))^2 $
in $V_C$ by an expectation value of order $a^2$, to replace $\rho^2$ by an
expectation value of order $(R/(MK_F))^{1/2} \sim a^2(R/a)^{1/2}$, and then to
multiply the result by the probability that the gluon is within a distance $a$
of $z= \pm R/2$.  This probability is of order $a/R$ at large $R$, as one sees
from the matching of the wave functions given in Eq. (\ref{match}).    Of
course, when $z\ll R$ the Coulomb potential is of order $\alpha_c /R$.  The
upshot is that all terms in the Coulomb potential are of this order:
\begin{equation}
\label{coulest}
V_C \simeq \frac{-A_1\alpha_c}{R[1+(R/a)^{1/2}]^{1/2}}- \frac{A_2\alpha_c}
{[R^2 +
a^2(R/a)^{1/2} ]^{1/2}}+\frac{\alpha_c}{R}.
\end{equation}
 where $A_{1,2}$ are positive coefficients of order unity.  Since we are not
treating the mesonic hybrid in detail, we have neglected
to write some other factors of order one at various other places in this
equation.

\end{document}